\documentclass[pra,twocolumn,a4paper]{revtex4}
\usepackage{amssymb}
\usepackage{epsfig}
\usepackage{amsmath}
\usepackage{graphicx}
\usepackage{bm}
\usepackage{times}
\usepackage{color}

\begin{document}

\title{ Micro-mechanical oscillator ground state cooling via intracavity
optical atomic excitations}
\author{C. Genes and H. Ritsch}
\affiliation{Institute for Theoretical Physics, University of Innsbruck, and Institute
for Quantum Optics and Quantum Information, Austrian Academy of Sciences,
Technikerstrasse 25, A-6020 Innsbruck, Austria}
\author{D. Vitali}
\affiliation{Dipartimento di Fisica, Universit\`{a} di Camerino, I-62032 Camerino (MC),
Italy}
\date{\today }

\begin{abstract}
We predict ground state cooling of a micro-mechanical oscillator, i.e. a vibrating end-mirror of an optical cavity, by resonant coupling of
mirror vibrations to a narrow internal optical transition of an ensemble of two level systems. The particles represented by a collective
mesoscopic spin model implement, together with the cavity, an efficient, frequency tailorable zero temperature loss channel which can be turned
to a gain channel of pump. As opposed to the case of resolved-sideband cavity cooling requiring a small cavity linewidth, one can work here with
low finesses and very small cavity volumes to enhance the light mirror and light atom coupling. The tailored loss and gain channels provide for
efficient removal of vibrational quanta and suppress reheating. In a simple physical picture of sideband cooling, the atoms shape the cavity
profile to enhance/inhibit scattering into higher/lower energy sidebands. The method should be applicable to other cavity based cooling schemes
for atomic and molecular gases as for molecular ensembles coupled to stripline cavities.
%(?? Tailorable reservoir response by help of inhomogeneously broadened
%ensembles). %David: why inhomogeneously broadened ?

%David: Stripline cavities coupled to molecular ensembles ?
\end{abstract}

\maketitle

Cavity cooling of micro-mechanical oscillators has seen tremendous
experimental progress
\cite{karrai,arcizet06b,bouwm,mavalvala,harris,lehnert08,Schliesser2009,Groeblacher2009}
in the past few years, also stimulated by theoretical analysis
showing that the quantum ground state of such oscillators is
achievable
\cite{Wilson-Rae07cooling,Marquardt07cooling,genes08cooling}. In
essence, through cavity mediated interaction the mechanical
vibrations at kHz-MHz frequencies are converted to strongly damped
sidebands of light modes at optical frequencies, that contain
essentially no thermal excitations (thermal photons). Standard
sideband cooling is effective only when the linewidth of the optical
resonator $\kappa$ is smaller than the mirror vibration frequency
$\omega _{m}$ and one has a sufficient optomechanical coupling. For
a given mirror reflectance a small enough $\kappa$ can be reached
using longer cavities, but strong optomechanical coupling then
requires a very large intracavity field and input power.

Here we propose the alternative use of atomic, molecular or solid
state electronic transitions as extremely localized narrow bandwidth
oscillators, for the cooling process. These are coupled to the
vibrational modes via a cavity light mode. Strong mirror-light and
light-atom coupling can be achieved while still maintaining a narrow
and tailorable resonance behavior. As the simplest generic example
we will consider ensembles of two-level atoms with a narrow
transition suitably detuned from the cavity resonance. At optical
frequencies thermal excitations of the atoms can be neglected and
the atoms constitute an effective zero temperature reservoir, where
energy and entropy can be dissipated efficiently. Note that using
optical resonances to cool solids has already been shown successful
in optical fibers~\cite{Rayner01}.
% relation to fiber cooling (Laser cooling of a solid from ambient temperature Authors: Rayner, A.1; Friese, M. E. J.1; Truscott, A. G.1;
% Heckenberg, N. R.1; Rubinsztein-Dunlop, H.1
% Journal of Modern Optics, Volume 48, Number 1, January 20, 2001 , pp. 103-114(12)

\begin{figure}[t]
\centerline{\includegraphics[width=0.45\textwidth]{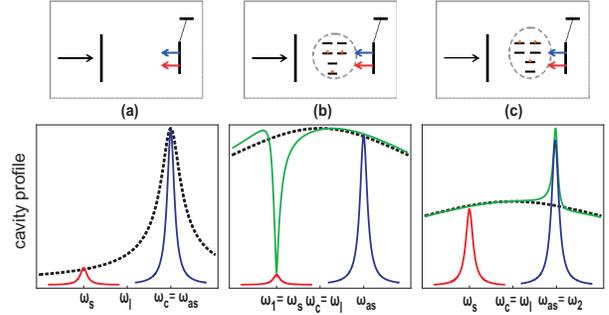}} \caption{(a) Typical resolved-sideband cooling of a micro-mirror in the
good-cavity regime is achieved via cavity enhancement of the blue (anti-Stokes) sideband. Laser driving is at frequency $\protect\omega _{l}$
while sidebands are scattered at $\protect\omega _{s}$ and $\protect\omega %
_{as}$. The dashed black line shows the cavity profile around the $\protect%
\omega _{c}$ cavity resonance. (b) Cooling via two-level atoms in the bad-cavity regime. The empty cavity response around $\protect\omega_{c}$
(dashed black line) is modified by the presence of the atoms of frequency splitting $\protect\omega _{1}$ which leads to a dip at
$\protect\omega _{s}$ (green line), thus effectively inhibiting Stokes scattering (heating). (c) Cooling via inverted two-level atoms with
splitting $\protect\omega _{2}$. The peak at $\protect\omega _{as}$ (green line) leads to enhancement of the anti-Stokes scattering (cooling).}
\label{scheme}
\end{figure}

Cavity assisted sideband cooling is based on unequal scattering of pump photons into higher/lower energy (anti-Stokes/Stokes) photons; the
cavity modifies the density of optical modes around the laser frequency such that a higher density at the anti-Stokes frequency leads to a
higher scattering rate into the more energetic sideband. Denoting the corresponding rates with $A_{as}$ and $A_{s}$, an effective extraction
rate of vibrational quanta $\Gamma =A_{as}-A_{s}$ can be added to the intrinsic
decay rate of the mechanical oscillator $\gamma _{m}$ to yield $\bar{\gamma}%
_{m}=\Gamma +\gamma _{m}\gg \gamma _{m}$. For an oscillator of frequency $%
\omega _{m}$, this produces cooling to an effective temperature $\bar{T}%
\simeq \hbar \omega _{m}/\left[ k_{B}\ln (1+1/\bar{n})\right] $ ($k_{B}=$%
Boltzmann's constant) where $\bar{n}=\gamma
_{m}n/\bar{\gamma}_{m}+n_{res}.$ Therefore cooling leads to a
reduction of the initial
average thermal vibration quantum number $n$ by a factor $\gamma _{m}/\bar{\gamma}_{m}$, %
%and is obviously useful only when one is able to start
%from a temperature low enough such that $\gamma _{m}n/\bar{\gamma}_{m}\ll 1$. However, even when
%this condition is fulfilled,
but one has also an additional residual occupancy $%
n^{res}=A_{s}/\bar{\gamma}_{m}$ that is due to the residual Stokes
scattering rate.
%amplitude for the ground state.
%One simple application of this principle is shown in \cite{karrai08dopplercooling} where a photonic
%crystal mirror near its photonic band gaps is used such that its
%reflectivity dependence with the wavelength has a large negative gradient.
Hence one best uses a high-finesse optical cavity (with %a small
$\kappa <\omega _{m}$, as illustrated in Fig.\thinspace \ref{scheme}a) resonant with the blue anti-Stokes sideband
\cite{Wilson-Rae07cooling,Marquardt07cooling} leading to cooling to $n^{res}\simeq \left( \kappa /2\omega _{m}\right) ^{2}$.

The novel mechanism that we propose in this Letter can be well understood from the picture of cavity cooling shown in Fig.\thinspace
\ref{scheme}a and is illustrated in Fig.\thinspace \ref{scheme}b,c. The added ensembles of two-level atoms, one prepared in the ground state
(type 1), and the other one inverted (type 2), modify the cavity susceptibility, and consequently change the cavity induced scattering to the
sidebands. Even in the limit of a cavity bandwidth much larger than the mirror vibration frequency the atoms still change the response function
to inhibit Stokes scattering and enhance scattering into the anti-Stokes sideband. As we will see later, the first mechanism is much more
efficient than the second in that, while it is characterized by slower cooling rates, it ensures a low $n^{res}\ll 1$, that allows for ground
state cooling. The difference comes from the fact that while the cavity filled with ground state atoms forms an effective zero-temperature bath,
the noise inherent in any amplification process implies an effective non-zero temperature near the gain maximum of the inverted atoms. Cooling
in the second case is then bounded by the effective bath temperature.

\textit{Hamiltonian} - A Fabry-Perot optical cavity is driven by a laser of frequency $\omega _{l}$ near one of the cavity resonances $\omega
_{c}$ with detuning $\Delta _{0}=\omega _{c}-\omega _{l}$. One of the cavity mirrors is allowed to vibrate at the mechanical frequency $\omega
_{m}$ and with an effective vibrating mass $m$. Two clouds of $N_{1,2}$ two-level atoms (TLA) with level splittings $\omega _{1,2}$ (detuned
from the laser by $\Delta _{1,2}=\omega _{1,2}-\omega _{l}$) are positioned inside the cavity mode. The Hamiltonian of the system is ($\hbar
=1$)
\begin{subequations}
\label{ham}
\begin{eqnarray}
H &=&H_{0}+H_{I}+H_{dis}, \\
H_{0} &=&\omega _{c}a^{\dagger }a+\frac{\omega _{m}}{2}\left(
p^{2}+q^{2}\right) +\omega _{1}S_{1z}+\omega _{2}S_{2z} \\
H_{I} &=&g_{1}\left( S_{1+}a+S_{1-}a^{\dagger }\right) +g_{2}\left(
S_{2+}a+S_{2-}a^{\dagger }\right) \\
&&-G_{0}a^{\dagger }aq+i\mathcal{E}\left( a^{\dagger }e^{-i\omega
_{l}t}-ae^{i\omega _{l}t}\right) ,  \notag
\end{eqnarray}%
where $H_{0}$ is the free part, $H_{I}$ is the interaction part while $%
H_{dis}$ is the dissipative part which will not be explicitly shown here but included in the Langevin equations. The free part includes the
cavity mode, mirror and atom free energy where $a$ is the annihilation field operator with $[ a,a^{\dagger }] =1$, $q$ and $p$ are the mirror
quadratures with $[ q,p] =i$ and the atoms are described by a set of collective spin operators satisfying $[ S_{1,2+},S_{1,2-}] =S_{1,2z}$ and
$[ S_{1,2z},S_{1,2\pm }] =\pm S_{1,2\pm }$. The interaction contains the atom-field Tavis-Cummings term and the radiation pressure field-mirror
coupling where $g_{1,2}$ is the single atom-field coupling strength while $G_{0}=(\omega_c/L)\sqrt{\hbar/m\omega_m}$ ($L$ is the cavity length)
is the single photon-mirror optomechanical coupling. The laser driving shows up in $H_{I}$ as a displacement term of amplitude $\mathcal{E}$
(directly connected to the input laser power). Not included in the Hamiltonian is a driving term that leads to an effective population inversion
in type 2 atoms.

\textit{Linearization} - For simplicity we perform a transformation from the
spin algebra to a harmonic oscillator algebra for the atomic system via $%
c_{1}=S_{1-}/\sqrt{N_{1}}$ and $c_{2}=S_{2+}/\sqrt{N_{2}}$, such that: $%
\omega _{1}S_{1z}\simeq \omega _{1}( -N_{1}/2+c_{1}^{\dagger }c_{1}) $ and $%
\omega _{2}S_{2z}\simeq \omega _{2}( N_{2}/2-c_{2}^{\dagger }c_{2})
$ and we have $[ c_{i},c_{j}^{\dagger }] =\delta_{ij}$. The effect
of $H_{dis}$ in Eq.~(\ref{ham}) can now be added in a Langevin
equation approach via the quantum noises $a_{in}$, $p_{in}$,
$c_{1,in}$ and $c_{2,in}$ with the following nontrivial
correlations: $\langle a_{in}( t) a_{in}^{\dagger }( t^{\prime })
\rangle =2\kappa \delta ( t-t^{\prime }) $, $\langle p_{in}( t)
,p_{in}( t^{\prime }) \rangle =\gamma _{m}( 2n+1) \delta (
t-t^{\prime }) $, $\langle c_{1,in}( t) c_{1,in}^{\dagger }(
t^{\prime }) \rangle =2\gamma _{1}\delta ( t-t^{\prime }) $ and
$\langle c_{2,in}( t) c_{2,in}^{\dagger }( t^{\prime })\rangle
=2\gamma _{2}\delta ( t-t^{\prime }) $ \cite{gard}. The cavity and
atom reservoirs are at zero-temperature and are fully characterized
by the decay rates $\kappa $, $\gamma _{1}$ and $\gamma _{2}$ while
the thermal mirror reservoir is described by the damping $\gamma
_{m}$ and thermal occupancy $n$. Notice that $\gamma _{1}$ is the
actual decay rate of atoms of type $1$ to their real ground state;
$\gamma_2$ is instead the effective optical pumping rate to the
excited state and can be considered as an effective ``decay'' rate.
The other levels involved in the optical pumping process are
irrelevant for cooling and are neglected in the present two-level
description.

As in \cite{genes08atoms}, one can linearize the equations of motion for the
atom-field-mirror system around steady state values to derive a set of
quantum linearized Langevin equations (QLLEs)
\end{subequations}
\begin{subequations}
\label{langlinear}
\begin{align}
\dot{q}& =\omega _{m}p, \\
\dot{p}& =-\omega _{m}q-\gamma _{m}p+G\left( a+a^{\dagger }\right) +p_{in},
\\
\dot{a}& =-\left( \kappa +i\Delta _{f}\right)
a+iGq-iG_{1}c_{1}-iG_{2}c_{2}^{\dagger }+a_{in}, \\
\dot{c}_{1}& =-\left( \gamma _{1}+i\Delta _{1}\right) c_{1}-iG_{1}a+c_{1,in},
\\
\dot{c}_{2}& =-\left( \gamma _{2}-i\Delta _{2}\right) c_{2}-iG_{2}a^{\dagger
}+c_{2,in}.
\end{align}%
where all the couplings are \textit{collectively enhanced},
$G_{1,2}=g_{1,2}\sqrt{N_{1,2}}$ and $G=G_{0}\alpha $, with $\alpha $
the steady state intracavity field amplitude (which can always be
assumed real).
%, $q_{s}$ for the mirror static position shift and $c_{s1}$ and $c_{s2}$ for the atoms.
In Eqs.~(\ref{langlinear}) $a\,$, $q$, $p$ and $c_{1,2}$ operators
denote steady state fluctuations, while
$\Delta_f=\Delta_0-G^2/\omega_m$ is the effective cavity detuning.

In the following it will be useful to define $\epsilon _{m}\left( \omega
\right) =\omega _{m}/\left( \omega _{m}^{2}-\omega ^{2}-i\gamma _{m}\omega
\right) $, $\epsilon _{f}\left( \omega \right) =1/\left[ \kappa +i\left(
\Delta _{f}-\omega \right) \right] $, $\epsilon _{1}\left( \omega \right) =1/%
\left[ \gamma _{1}+i\left( \Delta _{1}-\omega \right) \right] $ and $%
\epsilon _{2}\left( \omega \right) =1/\left[ \gamma _{2}-i\left(
\Delta _{2}+\omega \right) \right] $ as the mirror, field and atom
bare susceptibilities. Fourier transforming Eqs.~(\ref{langlinear})
one can derive the following set of coupled equations
\end{subequations}
\begin{subequations}
\label{coupled}
\begin{align}
\left[ \epsilon _{m}\left( \omega \right) \right] ^{-1}q& =G\left(
a+a^{\dagger }\right) +p_{in}, \\
\left[ \bar{\epsilon}_{f}\left( \omega \right) \right] ^{-1}a& =iGq+\bar{a}%
_{in},
\end{align}%
where the effective field input noise is $\bar{a}_{in}=a_{in}-i[
G_{1}\epsilon _{1}( \omega t) c_{1,in}+G_{2}\epsilon _{2}^{\ast }(
-\omega ) c_{2,in}^{\dagger }] $. The field response function is
modified by the atoms according to
\end{subequations}
\begin{equation}
\left[ \bar{\epsilon}_{f}\left( \omega \right) \right] ^{-1}=\left[ \epsilon
_{f}\left( \omega \right) \right] ^{-1}+G_{1}^{2}\epsilon _{1}\left( \omega
\right) -G_{2}^{2}\epsilon _{2}^{\ast }\left( -\omega \right) .
\label{response}
\end{equation}

\textit{Physical picture. Cavity response - }In the absence of atoms (setting $G_{1,2}=0$) Eqs.~(\ref{coupled}) describe a typical
optomechanical setup, which in the good-cavity limit $\kappa \ll \omega _{m}$ (resolved-sideband limit) and $\Delta _{f}=\omega _{m}$ leads to
optimal cooling of the mirror state. This effect can be simply understood in terms of preferential scattering of sidebands at frequencies
$\omega _{as,s}=\omega _{l}\pm \omega _{m}$ off the mirror, as illustrated in Fig.\thinspace \ref{scheme}a, where the cavity response (i.e.,
$\left\vert \epsilon _{f}\left( \omega \right) \right\vert $ is plotted around $\omega _{c}$) to the mechanical sidebands is depicted. The
choice of $\omega _{c}=\omega _{as}$ allows the enhancement of the cooling sideband. When atoms are placed inside the cavity, their effect on
the cavity response function leads to a modification of the cavity mode structure around the resonance $\omega _{c}$ quantified by Eq.\thinspace
(\ref{response}). This is illustrated in Fig.\thinspace \ref{scheme}b for the bad-cavity limit where $\kappa \gg \omega _{m}$, and resonant
cavity driving $\omega _{l}=\omega _{c}$. The type $1$ atoms, placed at $\Delta _{1}=-\omega _{m}$ induce a dip in the cavity profile at $\omega
_{s}$ while the gain medium increases the cavity response at $\omega _{as}$. What results is an inhibition of scattering processes into the
Stokes sideband and enhancement of scattering into the cooling sideband.

Based on the mode structure tailoring picture, we can now make some
quantitative remarks on the efficiency of cooling that is expected from such
a scheme. With the usual definition of the atom cooperativity $%
C_{1,2}=G_{1,2}^{2}/\left( \kappa \gamma _{1,2}\right) $, one can see that
the dip value\ at $\omega _{s}$ scales down as $\left( 1+C_{1}\right) ^{-1}$
while the peak at $\omega _{as}$ has a maximum that scales up $\left(
1-C_{2}\right) ^{-1}$. The widths of the dip/peak are $\bar{\gamma}%
_{1}=\gamma _{1}\left( 1+C_{1}\right) $ and $\bar{\gamma}_{2}=\gamma
_{2}\left( 1-C_{2}\right) $, respectively representing the
enhanced/reduced light -induced atomic linewidths. In the limit
where the scattered sidebands fit inside the dip/peak, i.e. the
resulting cooling rate is smaller than
$\bar{\gamma}_{1},\bar{\gamma}_{2}$, one then expects an inhibition
of the Stokes peak (enhancement of the anti-Stokes peak) by a factor
of the order of $\left( 1+C_{1}\right) ^{-1}$ ($\left(
1-C_{2}\right) ^{-1}$, respectively).

\textit{Cooling - }For a proper derivation of cooling rates and residual occupancy we start with Eqs. \thinspace (\ref{coupled}). Notice the
similarity with typical cavity-assisted cooling of a mirror, which can be obtained as a limiting case when $G_{1,2}=0$ for which
$\bar{a}_{in}=a_{in}$ and $\bar{\epsilon}_{f}\left( \omega \right) =\epsilon _{f}\left( \omega \right) $
\cite{Marquardt07cooling,genes08cooling}. The mechanical oscillator of Eq. \thinspace (\ref{coupled}a) is driven by a thermal zero-average
Langevin force $p_{in}$, and by a field-atom induced Langevin force $F=G\left( \bar{a}_{in}+\bar{a}_{in}^{\dagger }\right) $. Following the
approach of \cite{Marquardt07cooling}, one can find the cooling rate induced by the field-atom system and the mirror occupancy directly from the
spectrum of the force $F$, $\left\langle F\left( \omega \right) F\left( \omega ^{\prime }\right) \right\rangle =S_{F}\left( \omega \right)
\delta
\left( \omega +\omega ^{\prime }\right) $. From Eq. \thinspace (\ref{coupled}%
b) one derives
\begin{multline}
S_{F}\left( \omega \right) =2G\{\left\vert \bar{\epsilon}_{f}\left( \omega
\right) \right\vert ^{2}\left( \kappa +\gamma _{1}G_{1}^{2}\left\vert
\epsilon _{1}\left( \omega \right) \right\vert ^{2}\right) \\
+\left\vert \bar{\epsilon}_{f}\left( -\omega \right) \right\vert ^{2}\gamma
_{2}G_{2}^{2}\left\vert \epsilon _{2}\left( \omega \right) \right\vert
^{2}\}.  \label{spectrumforce}
\end{multline}
The cooling and heating rates (anti-Stokes and Stokes, respectively)
are computed as $A_{as,s}=$ $S_{F}\left( \pm \omega _{m}\right) /2$.
We shall consider from now $\Delta_f=0$, i.e., the cavity resonant
with the laser, so that there is no cooling in the absence of atoms.

\textit{Cooling with ground state atoms - }We restrict our discussion in the
following to a cavity filled with atoms of type 1 (setting $G_{2}=0$) and
derive the effective cooling rate and residual occupancy under resonance
condition $\Delta _{1}=-\omega _{m}$. First we stress that the effective
temperature of the cavity is zero: $\langle \bar{a}_{in}^{\dagger }(\omega )%
\bar{a}_{in}(\omega ^{\prime })\rangle =0$. We identify the optimal regime
as the bad-cavity regime $\kappa \gg \omega _{m}$, where the effect of the
atoms on the cavity profile is to efficiently inhibit Stokes scattering for
a bandwidth $\bar{\gamma}_{1}$ around $\omega _{s}$. We ask that the
bandwidth $\bar{\gamma}_{1}$ be larger than the scattered sideband
halfwidth, which is given by the total mechanical damping rate $\Gamma
_{1}+\gamma _{m}.$ In this limit and with the requirement $\gamma _{1}\ll
\omega _{m}$,. one can approximate $\left\vert \epsilon _{1}\left( \omega
_{m}\right) \right\vert \simeq 1/\left( 2\omega _{m}\right) $, $\left\vert
\epsilon _{1}\left( -\omega _{m}\right) \right\vert =1/\gamma _{1}$, $%
\left\vert \bar{\epsilon}_{f}\left( \omega _{m}\right) \right\vert \simeq
1/\kappa $ and $\left\vert \bar{\epsilon}_{f}\left( -\omega _{m}\right)
\right\vert \simeq 1/\left[ \kappa \left( 1+C_{1}\right) \right] $. The
scattering rates can be derived from Eq. (\ref{spectrumforce}) as $%
A_{as}\simeq G^{2}/\kappa $ and\ $A_{s}\simeq G^{2}/\left[ \kappa \left(
1+C_{1}\right) \right] $ and consequently the cooling rate is
\begin{equation}
\Gamma _{1}\simeq \frac{G^{2}}{\kappa }\frac{C_{1}}{1+C_{1}}.  \label{gamma1}
\end{equation}

\begin{figure}[tb]
\centerline{\includegraphics[width=0.45\textwidth]{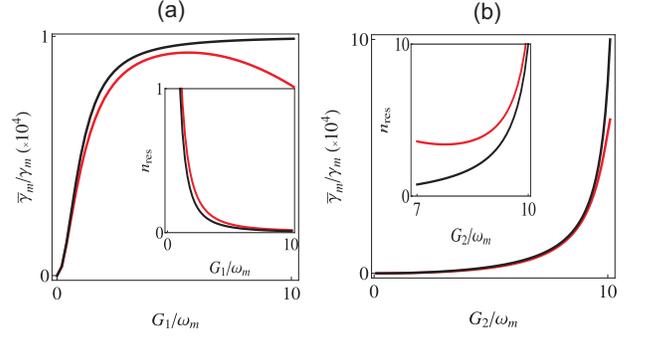}}
\caption{{\protect\small {}Relative cooling rate }$\bar{\protect\gamma}_{m}/$%
{\protect\small \ }$\protect\gamma _{m}${\protect\small \ and residual
occupancy }$n_{res}${\protect\small \ (in the insets) as a function of \ the
normalized parameters }$G_{1}/\protect\omega _{m}${\protect\small \ and }$%
G_{2}/\protect\omega _{m}${\protect\small \ for }$G=\protect\omega _{m}$%
{\protect\small . The red line corresponds to analytical expressions while the black line shows the exact results (see text). a) Cooling with
ground state atoms
and }$G_{2}=0${\protect\small . b) Cooling with inverted atoms and }$G_{1}=0$%
{\protect\small . The other parameters are }$\protect\omega _{m}/2\protect%
\pi =10${\protect\small \ MHz, }$\protect\gamma _{m}=10^{-5}\times \protect%
\omega _{m}${\protect\small , }$\protect\kappa =100\times \protect\omega %
_{m} ${\protect\small , }$\protect\gamma _{1}=\protect\omega _{m}/100$%
{\protect\small , }$\protect\gamma _{2}=\protect\omega _{m}${\protect\small %
\ }$\ ${\protect\small and }$n=100.$}
\label{coolingoccupancy}
\end{figure}

The residual occupancy can be computed as $n_{1}^{res}=A_{s}/\bar{\gamma}%
_{m}\simeq A_{s}/\Gamma _{1}\rightarrow C_{1}^{-1}$ (in the limit of large cooperativity). This result replaces the resolved-sideband limit
final occupancy $n^{res}=\left( \kappa /2\omega _{m}\right) ^{2}$ of the purely optomechanical case with no atoms. The advantage of adding
ground state atoms within the cavity is obvious from the scaling of $n_{1}^{res}$ with the controllable and in principle unbounded parameter
$N_{1}$. In fact, in the strong resolved-sideband limit $\kappa \ll \omega _{m}$, the cavity is far-off resonance, and one has to use large
input power for optimal cooling, while with atoms one can make $n_{1}^{res}$ smaller and smaller by increasing $N_{1}$. Illustrations of the
dependence of $\bar{\gamma}_{m}/\gamma _{m}$ and $n_{1}^{res}$ on the normalized coupling strengths $G_{1}/\omega _{m}$ when $G_{2}=0$ is shown
in Fig.~\ref{coolingoccupancy}a where red lines correspond to the exact numerical solution of Eqs. (\ref{langlinear}) in steady state to derive
the $4$-mode covariance matrix, while black lines show the corresponding previously derived analytical expressions. The set of parameters we
have made use of is $\omega _{m}/2\pi =10$ MHz, $\gamma _{m}=10^{-5}\times \omega _{m}$, $\kappa =100\times \omega _{m}$, $\gamma _{1}=\omega
_{m}/100$ and $n=100$.

The condition $\Gamma _{1}+\gamma _{m}<\bar{\gamma}_{2}$ puts a lower bound
on $G_{1}$ which is given by $G$. Moreover, with the increase of $G_{1}$,
the field response at $\omega _{as}$ ($\left\vert \bar{\epsilon}_{f}\left(
\omega _{m}\right) \right\vert ^{2}$) diminishes as $1+\left(
G_{1}^{2}/2\kappa \omega _{m}\right) ^{2}$ from the value of $1/\kappa ^{2}$
assumed above, which in consequence amounts to the limitation that $G_{1}<%
\sqrt{\kappa \omega _{m}}$. The cooling rate is independent of $G_{1}$ as
long as $G_{1}>G$, however the residual occupancy is limited by $%
n_{1}^{res}\simeq C_{1}^{-1}>\gamma _{1}/\omega _{m}$, which is the
justification for working with small $\gamma _{1}.$

\textit{Cooling with inverted atoms - }We shift our discussion now to the
opposite case of a cavity filled with atoms of type $2$, at $\Delta
_{2}=\omega _{m}$ and set $G_{1}=0$. The correlations of the input noises
are now $\langle \bar{a}_{in}^{\dagger }(\omega )\bar{a}_{in}(\omega
^{\prime })\rangle =2\gamma _{2}G_{2}^{2}|\epsilon _{2}(\omega )|^{2}\delta
(\omega +\omega ^{\prime })$ and $\langle \bar{a}_{in}(\omega )\bar{a}%
_{in}^{\dagger }(\omega ^{\prime })\rangle =2\kappa \delta (\omega +\omega
^{\prime })$ which show that the cavity is at a nonzero effective
temperature. This results in a frequency-dependent thermal cavity occupancy $%
n\left( \omega \right) =\gamma _{2}G_{2}^{2}|\epsilon _{2}(\omega )|^{2}\kappa ^{-1}(1-\gamma _{2}G_{2}^{2}|\epsilon _{2}(\omega )|^{2}/\kappa
)^{-1}$ which evaluated at $\omega _{m}$ leads to $n\left( \omega _{m}\right) =C_{2}/\left( 1-C_{2}\right) $. This will set a lower bound for
the achievable mechanical occupancy because the oscillator cannot be cooled to less than the occupancy of the effective cooling reservoir. In
the bad cavity limit the atoms produce a peak of bandwidth $\bar{\gamma}_{2}$ around $\omega _{as}$. We ask again for the condition that the
anti-Stokes sideband fits inside the peak, i.e. $\Gamma _{2}+\gamma _{m}<$ $\bar{\gamma}_{2}$. In this limit and with
the requirement $\gamma _{2}\ll \omega _{m}$, one can approximate $%
\left\vert \epsilon _{2}\left( \omega _{m}\right) \right\vert \simeq
1/\left( 2\omega _{m}\right) $, $\left\vert \epsilon _{2}\left( -\omega
_{m}\right) \right\vert =1/\gamma _{2}$, $\left\vert \bar{\epsilon}%
_{f}\left( \omega _{m}\right) \right\vert \simeq 1/\left[ \left(
1-C_{2}\right) \kappa \right] $ and $\left\vert \bar{\epsilon}_{f}\left(
-\omega _{m}\right) \right\vert \simeq 1/\kappa $. The scattering rates are $%
A_{as}\simeq G^{2}/\left[ \kappa \left( 1-C_{2}\right) ^{2}\right] $ and\ $%
A_{s}\simeq G^{2}/\kappa \left[ 1+C_{2}/\left( 1-C_{2}\right) ^{2}\right] $
and the cooling rate is
\begin{equation}
\Gamma _{2}\simeq \frac{G^{2}}{\kappa }\frac{C_{2}}{1-C_{2}}.  \label{gamma2}
\end{equation}%
The residual occupancy in the limit of $C_{2}\rightarrow 1$ diverges as $%
n_{2}^{res}\simeq C_{2}/\left( 1-C_{2}\right) $, which is the occupancy of
the effective reservoir at the Stokes sideband. However, $C_{2}$ cannot be
increased arbitrarily close to $1$ since there is a constraint $\Gamma
_{2}+\gamma _{m}<$ $\bar{\gamma}_{2}$ which leads to $G^{2}/\left( \gamma
_{2}\kappa \right) <\left( 1-C_{2}\right) ^{2}$. Graphical illustration of
the increase of the normalized cooling rate $\bar{\gamma}_{m}/\gamma _{m}$
as a function of $G_{2}/\omega _{m}$ at fixed $G_{1}=0$ is provided in Fig.~%
\ref{coolingoccupancy}b. The analytical result is valid up to values of $%
G_{2}$ where $\bar{\gamma}_{m}$ is of the order of $\bar{\gamma}_{2}$. The
inset shows the agreement between the analytical and numerical results for $%
n_{2}^{res}$ when $G_{2}$ is large such that $C_{2}\rightarrow 1$.

\textit{Discussion - }The use of ground state atoms for cooling
presents the advantage of a low residual occupancy proportional to
$C_{1}^{-1}$. The downside is that cooling is slow at a rate that
saturates at $G^{2}/\kappa $ in the limit $C_{1}\gg 1$. The use of
inverted atoms considerably speeds up the cooling process that takes
place at a rate $\Gamma _{2}\gg G^{2}/\kappa $ $\ $for $C_{2}$ close
to unity, at the price, however, of a high $n_{res}$. One would
ideally combine the two methods for optimal cooling. However, such a
mixed scenario is not particularly advantageous. In fact, in the
limit of $\gamma _{1,2}\ll \omega _{m}$, one obtains a combined
cooling rate that is practically the one induced by the anti-Stokes
enhancement $\Gamma _{12}\simeq $ $\Gamma _{2}$, but the residual
occupancy $n_{12}^{res}\simeq n_{1}^{res}(1-C_{2})+n_{2}^{res}\simeq
n_{2}^{res}$ is still large, reflecting the negative effect of the
high effective temperature induced by the inverted atoms.

A few remarks on the system stability have to be made at this point. The analysis can be made by applying the Routh-Hurwitz criterion
\cite{grad} on Eqs.~(\ref{langlinear}). A cavity with a moving mirror driven at resonance is always stable and its stability is not perturbed by
the addition of ground state atoms at $\Delta _{1}=-\omega _{m}$. The inverted atoms, however, do impose a limitation on the system's stability,
which can be
simply cast as $C_{2}<1$ and trivially fulfilled by the proper choice of $%
N_{2}$.

The question of the validity of bosonisation also has to be discussed. For
the ensemble of two level atoms to resemble an harmonic oscillator one has
to ask that the single atom excitation induced by the cavity field is small.
This means $g_{1,2}^{2}/\left( \omega _{m}^{2}+\gamma _{1,2}^{2}\right) \ll
\alpha ^{-2}\ll 1$, which, as detailed in \cite{genes08atoms} can be
fulfilled for a dipole forbidden transition or for atomic vapor cell much
larger than the cavity mode.

\textit{Conclusions - }We have shown a new mechanism for mirror ground state cooling where atoms are used to dissipate the thermal energy of the
mechanical oscillator. In a simple picture this can be seen as cavity-assisted cooling with a cavity profile tailored via use of atoms at the
Stokes and anti-Stokes sidebands. The effect is to inhibit scattering processes that lead to heating while enhancing the cooling process. We
notice that there are alternative ways of cooling through cavity spectrum tailoring. A recent example is given in \cite{Elste09} where tailoring
is obtained when the micro-mechanical resonator directly modulates the cavity bandwidth. More generally speaking, appropriate spectrum tailoring
can be always be obtained through quantum interference \cite{Rodrigues09}, as for example proposed and demonstrated for cooling trapped ions
\cite{Morigi03}.

\textit{Acknowledgements - } We are grateful to G. Milburn and M.
Wallquist for helpful discussions. We acknowledge financial support
from the European Commission (FP6 Integrated Project QAP, and
FET-Open project MINOS) and Euroquam Austrian Science Fund project
I119 N16 CMMC.


\begin{thebibliography}{9}

\bibitem{karrai} C. H. Metzger and K. Karrai, Nature (London), \textbf{432},
1002 (2004).

\bibitem{arcizet06b} O. Arcizet, P.-F. Cohadon, T. Briant, M. Pinard, and A.
Heidmann, Nature (London) \textbf{444}, 71 (2006).

\bibitem{bouwm} D. Kleckner and D. Bouwmeester, Nature (London) \textbf{444}%
, 75 (2006).

\bibitem{mavalvala} T. Corbitt, Y. Chen, E. Innerhofer, H. M\"uller-Ebhardt,
D. Ottaway, H. Rehbein, D. Sigg, S. Whitcomb, C. Wipf, and N. Mavalvala, Phys. Rev. Lett. \textbf{98}, 150802 (2007).

\bibitem{harris} J. D. Thompson, B. M. Zwickl, A. M. Jayich, F. Marquardt,
S. M. Girvin, and J. G. E. Harris, Nature (London) \textbf{452}, 72 (2008).

\bibitem{lehnert08}J. D. Teufel, J. W. Harlow, C. A. Regal, and K. W. Lehnert, Phys. Rev. Lett. \textbf{101}, 197203 (2008).

\bibitem{Schliesser2009} A. Schliesser, O. Arcizet, R. Rivi\`{e}re, T. J.
Kippenberg, arXiv:0901.1456v1

\bibitem{Groeblacher2009} S. Groeblacher, J.B. Hertzberg, M.R. Vanner, S.
Gigan, K.C. Schwab, M. Aspelmeyer, arXiv:0901.1801v1

%\bibitem{karrai08dopplercooling} K. Karrai, I. Favero and C. Metzger, Phys.
%Rev. Lett \textbf{100} 240801 (2008)

\bibitem{Wilson-Rae07cooling} I. Wilson-Rae, N. Nooshi, W. Zwerger, and T.
J. Kippenberg, Phys. Rev. Lett. \textbf{99}, 093901 (2007)

\bibitem{Marquardt07cooling} F. Marquardt, J.P. Chen, A.A. Clerk,
and S. M. Girvin Phys. Rev. Lett. \textbf{99}, 093902 (2007).

\bibitem{genes08cooling} C. Genes, D. Vitali, P. Tombesi, S. Gigan, and M.
Aspelmeyer, Phys. Rev. A \textbf{77}, 033804 (2008)

\bibitem{Rayner01}
% relation to fiber cooling (Laser cooling of a solid from ambient temperature Authors:
A. Rayner, M.E.J. Friese, A.G. Truscott, N. Heckenberg, and H. Rubinsztein-Dunlop, J. Mod. Opt. \textbf{48}, 103(2001).

\bibitem{gard}C. W. Gardiner and P. Zoller, \textit{Quantum Noise},
(Springer, Berlin, 2000).

\bibitem{genes08atoms} C. Genes, D. Vitali, and P. Tombesi, Phys. Rev. A
\textbf{77}, 050307 (2008).

\bibitem{grad} I. S. Gradshteyn and I. M. Ryzhik, Table of Integrals, Series
and Products, Academic Press, Orlando, pag. 1119, (1980).

\bibitem{Elste09}F. Elste, S.M. Girvin, and A.A. Clerk, Phys. Rev. Lett. \textbf{102}, 207209 (2009).

\bibitem{Rodrigues09}D.A. Rodrigues, Phys. Rev. Lett. \textbf{102}, 067202 (2009).

\bibitem{Morigi03}G. Morigi, Phys. Rev. A \textbf{67}, 033402 (2003); G. Morigi, J. Eschner, and C.H. Keitel, Phys. Rev. Lett. \textbf{85},
4458 (2000); C.F. Roos, D. Leibfried, A. Mundt, F. Schmidt-Kaler, J. Eschner, and R. Blatt, Phys. Rev. Lett. \textbf{85}, 5547 (2000).


\end{thebibliography}
\end{document}